\documentstyle[psfig]{l-aa}

\begin{document}

   \thesaurus{06         
              (03.11.1;  
               16.06.1;  
               19.06.1;  
               19.37.1;  
               19.53.1;  
               19.63.1)} 
   \title{Near IR photometry of the old open clusters Berkeley~17 and
Berkeley~18\thanks{Based on observations taken at TIRGO.}}

\subtitle{Probing the Age of the Galactic Disc}

   \author{Giovanni Carraro 
          \inst{1}, Antonella Vallenari\inst{2}, L\'eo Girardi\inst{3}
	  and Andrea Richichi\inst{4}
          }

   \offprints{Giovanni Carraro ({\tt carraro@pd.astro.it})}

   \institute{Dipartimento di Astronomia, Universit\'a di Padova,
	vicolo dell'Osservatorio 5, I-35122, Padova, Italy
        \and
        Osservatorio Astronomico di Padova, vicolo Osservatorio 
          5, I-35122, Padova,
	Italy
        \and
        Max Planck Institut f\"ur Astrophysik, Karl Schwarzschild Str. 1
        D-87548, Garching bei M\"unchen, Germany
        \and
        Osservatorio Astrofisico di Arcetri, Largo E. Fermi 5,
        I-50125, Firenze, Italy\\
        e-mail: {\tt 
carraro,vallenari\char64pd.astro.it,leo\char64mpa-garching.mpg.de,
richichi\char64arcetri.fi.astro.it}
             }

   \date{Received November 6; accepted December 15, 1998}

   \maketitle

   \markboth{Carraro et al}{Old Open Clusters}

   \begin{abstract}

We report on near IR ($J$ and $K$ bands) observations of two $8 \times 8
~(arcmin)^{2}$ regions centered on the old open clusters Berkeley~17 and
Berkeley~18, for which only optical photometry (in $B$,$V$ and $I$ bands)
exists. $J$ and $K$ photometry allows us to obtain an independent estimate
of cluster metallicity by means of the relationship between the
spectroscopic metallicity and the Red Giant Branch (RGB) slope calibrated
by Tiede et al (1997). \\
From the analysis of the colour magnitude diagram (CMD) and
luminosity function (LF), Berkeley~17 turns out to have a metal content 
$[Fe/H]~\sim~-0.35$.  It is $9~Gyr$ old, suffers from a
reddening $E(B-V) = 0.58~mag$ and has an heliocentric  distance of
$2.5~kpc$. Berkeley~17 comes out to be substantially younger than in
previous work (age $\approx 13~Gyr$).\\
On the other hand Berkeley~18 is found to have solar metal abundance, and
to be younger than Berkeley~17,
with an age of about $4~Gyr$. While we confirm Kaluzny (1997) reddening
estimate,
we significantly revise the distance of the cluster, which lies $4.5~kpc$
from the Sun. 
These results on two open clusters believed to be between the oldest put
constraints on the age
and the evolution of the Galactic Disc. The absence of clusters older than
$8-9~Gyr$ suggests the possibility that the Galaxy underwent a star
formation minimum between $13$ and $10~Gyr$ ago.

      \keywords{Galactic Disc: Formation and Evolution --
                Open clusters and associations: individual: 
                Berkeley~17 and Berkeley~18                
               }
   \end{abstract}

%

\section{Introduction}
Whether the Galaxy formed by an homogeneous monolithic collapse {\it a
l\'a} Eggen et al (1962), or by a series of merging and/or capture events 
{\it a l\'a} Searle \& Zinn (1978) is still an open question.
The traditional approach to investigate this fundamental issue is the
comparative study of different stellar populations long ago
recognized to exist inside the Milky Way (halo, disc - thin and thick -
and bulge).\\
In particular a decisive step is to understand  the formation and
evolution of the Galactic Disc, and its relation with the Halo and the
Bulge.
Open clusters, especially the oldest ones, are the best candidates to
trace the early stages of the disc  development, the chemical and
dynamical evolution of the disc, and the possible existence of a hiatus in
the star formation history of the Milky Way between the formation of the
halo/thick disc and the thin disc.\\
Janes \& Phelps (1994) discussed in details some of these interesting
questions, suggesting that :
\begin{description}
\item $\bullet$ the age distribution of the open clusters overlaps that of
globular clusters suggesting that the disc started to form before the end
of the halo formation;
\item $\bullet$ captures events were responsible for the formation of the
disc radial metallicity gradient;
\item $\bullet$ the overdensity of open clusters in the age  range
between 5-7 Gyr is  probably due to a burst of star formation induced
by a merging episode.
\end{description}

The first point is of particular interest. The conclusion of the authors
completely relies on the existence of a cluster, Berkeley~17, whose age
is estimated around 12 Gyr (Phelps 1997). 
The bulk of globular clusters has ages around 13 Gyr (Gratton et al
1997) with a tail down to $11~Gyr$. This opens the
possibility that there has been no hiatus at all between the formation
of the halo/thick disc and the thin disc.

\begin{table}
\tabcolsep 0.10truecm
\caption{Basic parameters of the studied clusters.}
\begin{tabular}{lccccc}
\hline
\hline
\multicolumn{1}{c}{Cluster} &
\multicolumn{1}{c}{$\alpha_{2000.0}$} &
\multicolumn{1}{c}{$\delta_{2000.0}$} &
\multicolumn{1}{c}{$l$} &
\multicolumn{1}{c}{$b$} &
\multicolumn{1}{c}{Diameter}\\
 &$hh~mm$& $^{o}$ &$^{o}$ &$^{o}$ &($\prime$) \\   
\hline
Berkeley~17    & 05~17.4 & 30:33 & 175.65 & -3.65 & 15\\
Berkeley~18    & 05~18.5 & 45:21 & 163.63 & +5.01 & 26\\
\hline\hline
\end{tabular}
\end{table}

\begin{table*}
\caption[ ]{ Observation Log Book }
\begin{tabular}{c|c|c|c|c|c}
\hline
\hline
Cluster               &$\alpha$    &$\delta$  & Date& \multicolumn{2}{c}
{Exposure Times (sec)} \\ 
                           &(2000)      &(2000) &    &J&K\\
\hline
Be~17&5 20 32& 33 34 40& Oct, 27, 1997& 840&844\\
Be~18& 05 22 07 & 45 25 17& Oct,25, 1997 & 1020  &960   \\
\hline
\hline
\end{tabular}
\end{table*}

In this paper we would like to address the issue of the age of
Berkeley~17, 
analyzing IR photometry in $J$ and $K$ bands. We  also discuss similar
data for 
Berkeley~18, another old open
cluster.  This is part of a project to study northern open clusters in the IR.
We report elsewhere (Vallenari et al 1998) on two very young open clusters,
Berkeley~86 and NGC~1893.\\
The age of Berkeley~17 has been discussed by Carraro
et al (1998) using published $BVI$ photometry. From this study it emerges
that Berkeley~17 is about 9 Gyr
old, and the oldest open cluster, NGC~6791, is 3-4 Gyr younger than the
bulk of globular clusters, suggesting the possibility that a real 
star formation minimum occurred between the halo and disc settling.\\
Berkeley~17 and Berkeley~18 were discovered by Setteducati \& Waever
(1962) in their search for unstudied open clusters.
Both the clusters were never observed in the IR, whereas they have been
studied in the optical ($B$, $V$ and $I$). Berkeley~17 was studied by
Kaluzny (1994) and Phelps(1997), whereas Berkeley~18 was studied by
Kaluzny (1997). Their basic properties are summarized in Table~1.\\
This paper is organized as follows: Section~2 presents the observation and
data reductions; Section~3 discusses 
CMDs of the clusters under investigation,
whereas Sections~4 and 5 deal with the derivation of the clusters' fundamental
parameters. Finally Section~6 gives some concluding remarks.

\begin{figure}
\centerline{\psfig{file=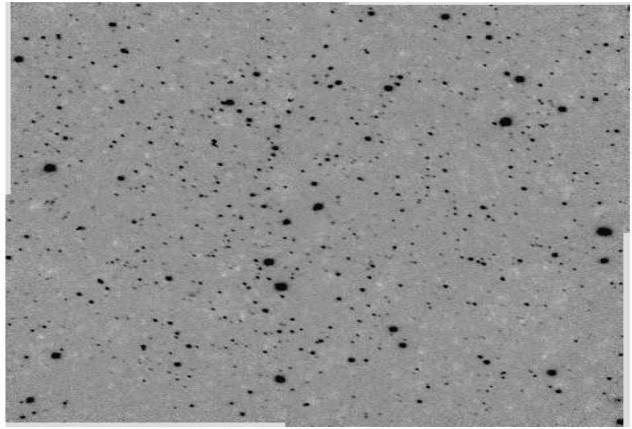,width=9cm,height=9cm}}
\caption{A mosaic of the four regions observed in the field of
Berkeley~17.}
\end{figure}

\begin{figure}
\centerline{\psfig{file=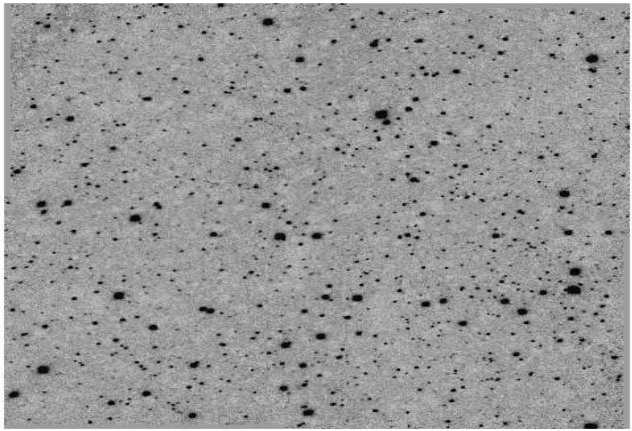,width=9cm,height=9cm}}
\caption{A mosaic of the four regions observed in the field of
Berkeley~18.}
\end{figure}

\section {Observations and Data reduction}

J (1.2 $\mu$m) and K (2.2 $\mu$m) photometry of the two clusters
was obtained 
at the  1.5m Gornergrat Infrared Telescope (TIRGO) 
equipped with Arcetri Near Infrared Camera (ARNICA)
in October 1997. 
ARNICA is based  on a NICMOS3 256$\times$ 256
pixels array (gain=20 e$^-/$ADU, read-out noise=50 e$^-$,
 angular scale =1$\arcsec/$pixel, and $4 \time 4 \arcmin^{2}$ field of
view). 
Through each filter 4 partially overlapping images of
each cluster were obtained,
 covering a total field of view of about 8 $\times 8 \arcmin^{2}$,
in short exposures to avoid sky saturation.
The two  fields  are
located close to the center of Berkeley~17 and Berkeley~18.
 The log-book of the observations is presented
 in Table~2 where the centers of the observed fields and the total
exposure times are given.
 The nights were photometric
with a seeing of 1$\arcsec$-1.5$\arcsec$. 
Fig.1a,b presents the final mosaics of the 4 frames
  for both clusters in K passband.

The data were reduced  subtracting from each image 
a linear combination of the
corresponding  skies and dividing the results by
the flat-field. We make use of the Arnica package (Hunt et al 1994)
in IRAF and Daophot II. 
The conversion of the
instrumental magnitude j and k to the standard
J, K was made using stellar fields of standard stars taken
from   Hunt et al (1998) list.
About 10 standard stars per night have been used.
The  relations in usage per 1 sec exposure time are: 
\begin{equation}
J  = j+19.51
\end{equation}
\begin{equation}
K  = k+18.94
\end{equation}
with standard deviation of the zero points of 0.03  mag for the $J$ 
and 0.04 for the $K$ magnitude. This error is only due to the linear
interpolation of the standard stars. The calibration uncertainty
is dominated by the error due to the correction from aperture photometry
to PSF fitting magnitude. 
The standard stars used for the calibration do  not cover the entire
colour range of the data, because of the lack of stars redder than
$(J-K) \sim 0.8$. From our data, no colour term is found for $K$ mag,
whereas we cannot exclude 
it  for the $J$ magnitude.
Taking all into account,
 we estimate that the total error
on the calibration is about 0.1 mag both in $J$ and $K$ pass-bands.

\begin{figure*}
\centerline{\psfig{file=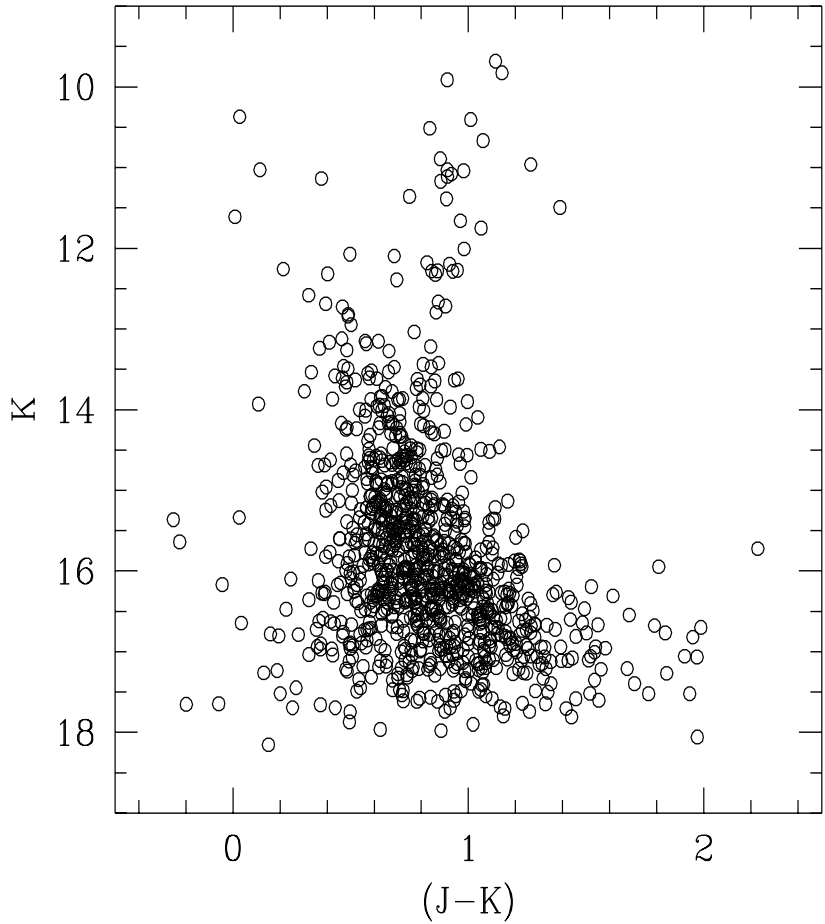,width=14cm,height=16cm}}
\caption{The CMD of Berkeley~17.}
\end{figure*}

\begin{figure*}
\centerline{\psfig{file=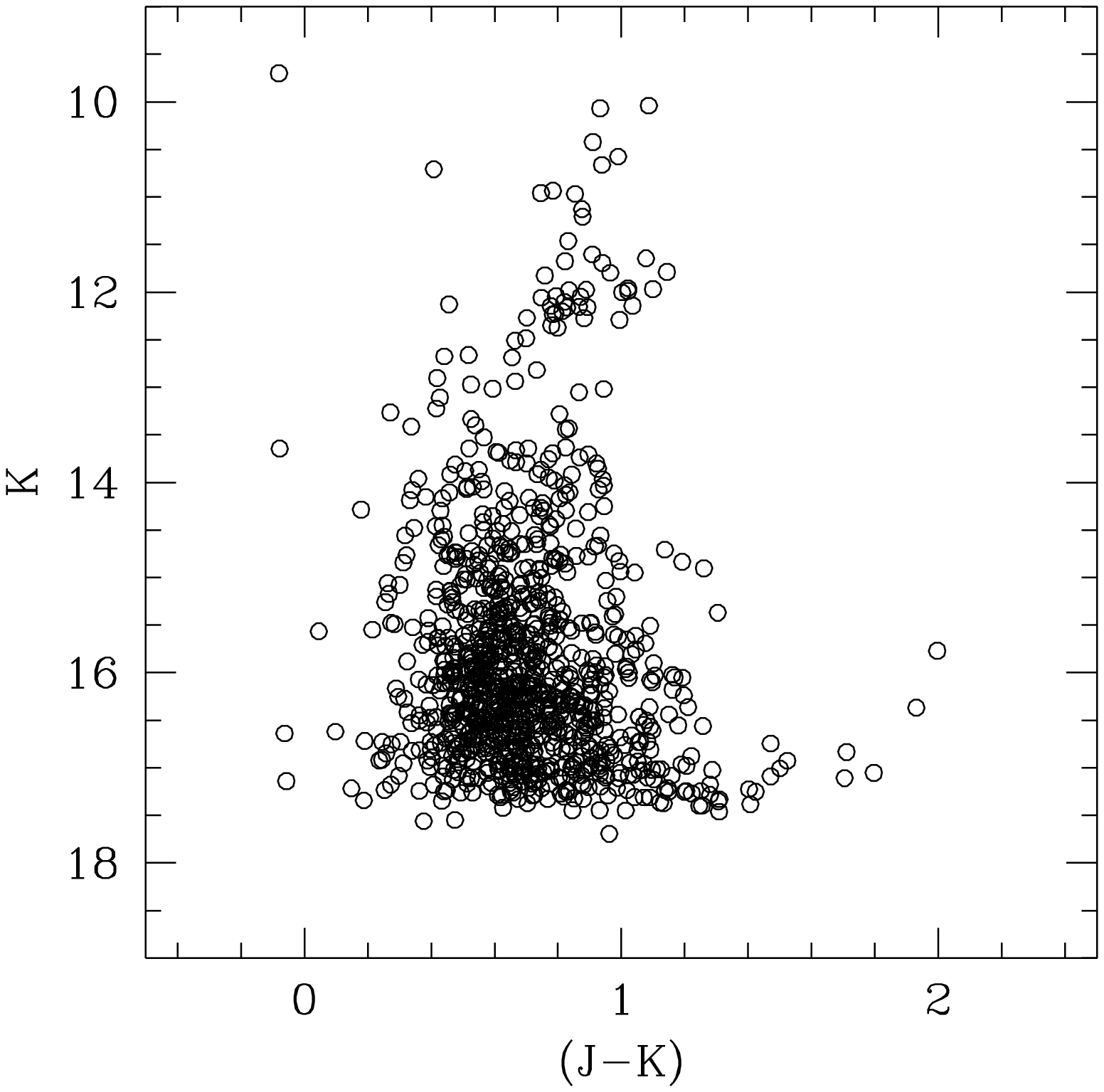,width=14cm,height=16cm}}
\caption{The CMD of Berkeley~18}
\end{figure*}

\section{The Color-Magnitude Diagrams}
$K-(J-K)$ relationship for Berkeley~17 and Berkeley~18 are shown in
Fig~3 and Fig~4, respectively.

\subsection{Berkeley~17}
The CMD of Berkeley~17 shows all the features of a very old open cluster,
namely a scarcely  populated extended Red Giant Branch (RGB) and a clump
of He-burning stars.
Some  field stars contamination from the galactic disc is visible in the
main sequence (MS), which extends almost vertically well above the cluster
Termination 
Point (TO) up to $K = 11$.\\
The cluster TO  is most probably located at $K \approx 14.5$ and
$(J-K) \approx  0.50$, whereas the  clump of He-burning stars is visible
at $K \approx 11.20$ and $(J-K) \approx 0.90$. It is composed  by 4-5
stars, which are the counterpart of the group of stars located at
$V \approx 14.5$, $(V-I)\approx 1.7$ in the $BVI$ photometry of Phelps
(1997).  The brighter part of our CMD is somewhat more populated than
Phelps (1997) study, because of the larger field coverage ($8^{\prime}
\times 8^{\prime}$ compared to $5.1^{\prime} \times 5.1^{\prime}$). 
As a consequence, the RGB is better defined in our study.\\
The widening of the MS is due to many causes (binaries, differential
reddening and so forth), the predominant one being the
photometric errors, which are 0.06, 0.12 and 0.17  at $K$ = 16, 17 and 18,
respectively.
The magnitude difference $\Delta K$ between the TO and the clump turns out
to be about $3.2~ mag$.

\subsection{Berkeley~18}
As for Berkeley~17, Berkeley~18 shows to be an old cluster.
It appears to be less contaminated in field foreground/background stars,
the MS turning off at $K \approx 15.00$ and $(J-K) \approx 0.40$.
The RGB  shows some gaps, whereas the clump of He-burners (about 10 stars)
is located at $K \approx 12.25$ and $(J-K) \approx 0.80$.
The magnitude difference $\Delta K$ turns out to be about $2.7$,
suggesting that Berkeley~18 is younger than Berkeley~17.
We notice that the CMD shows the presence of a group of stars above
the clump, which does not lie along the RGB. These are probably field
stars, which are also present  in the photometric $BVI$ survey of
Kaluzny
(1997). \\
Also in the case of Berkeley~18 we ascribe the MS widening to
binary and field stars contamination, possible differential reddening
and photometric errors. These latter are 0.08, 0.10, 0.18 at $K$ = 15, 
17 and 18, respectively.

\section{Berkeley~17: cluster parameters}

\subsection{Metallicity}
Friel et al (1995) measured Berkeley~17 metal abundance using moderate
resolution spectroscopy of 12 stars. They obtained $[Fe/H] =
-0.29\pm0.13$. It is possible to derive the cluster abundance
photometrically in the IR using the calibration of Tiede et al (1997).
This relation correlates the abundance $[Fe/H]$ with the slope
$\Delta(J-K) / \Delta K$ of rhe RGB.\\
It reads:

\begin{equation}
[Fe/H] = -2.98 (\pm0.70) -23.84(\pm6.83 \times (RGB slope)
\end{equation}

This relation has been proved to hold for both  metal rich globular
clusters and  open clusters. For the latter the RGB slope tends to
present  less negative values at decreasing age.
We applied this relation to Berkeley~17, for which the RGB slope is
$\Delta(J-K) / \Delta K = -0.11\pm 0.02$. This results in
$[Fe/H] \sim  -0.35$, consistent  with  spectroscopic estimates.

\subsection{Age, Distance and Reddening}
Berkeley~17 was studied in the optical by Kaluzny (1994) and Phelps
(1997).\\
Kaluzny(1994) compared Berkeley~17 with NGC~6791 and, assuming that the
former is metal poorer than the latter, concluded that the two clusters
are probably coeval (around $9~Gyr$ old). Moreover he suggested a  
distance modulus $(m-M) > 15.0~mag$ and a reddening $E(V-I) > 0.70~mag$.
On the other hand, Phelps (1997) reached considerably different results,
although the two photometric studied were not significantly different
(see Carraro et al 1998). In details he found an age of
$12^{+1}_{-2}~Gyr$, a distance modulus $(m-M) \approx 13.9~mag$ and a 
reddening in the range $0.52~<~E(B-V)~<~0.68~mag$ and 
$0.61~<~E(V-I)~<0.71~mag$.\\

\begin{figure}
\centerline{\psfig{file=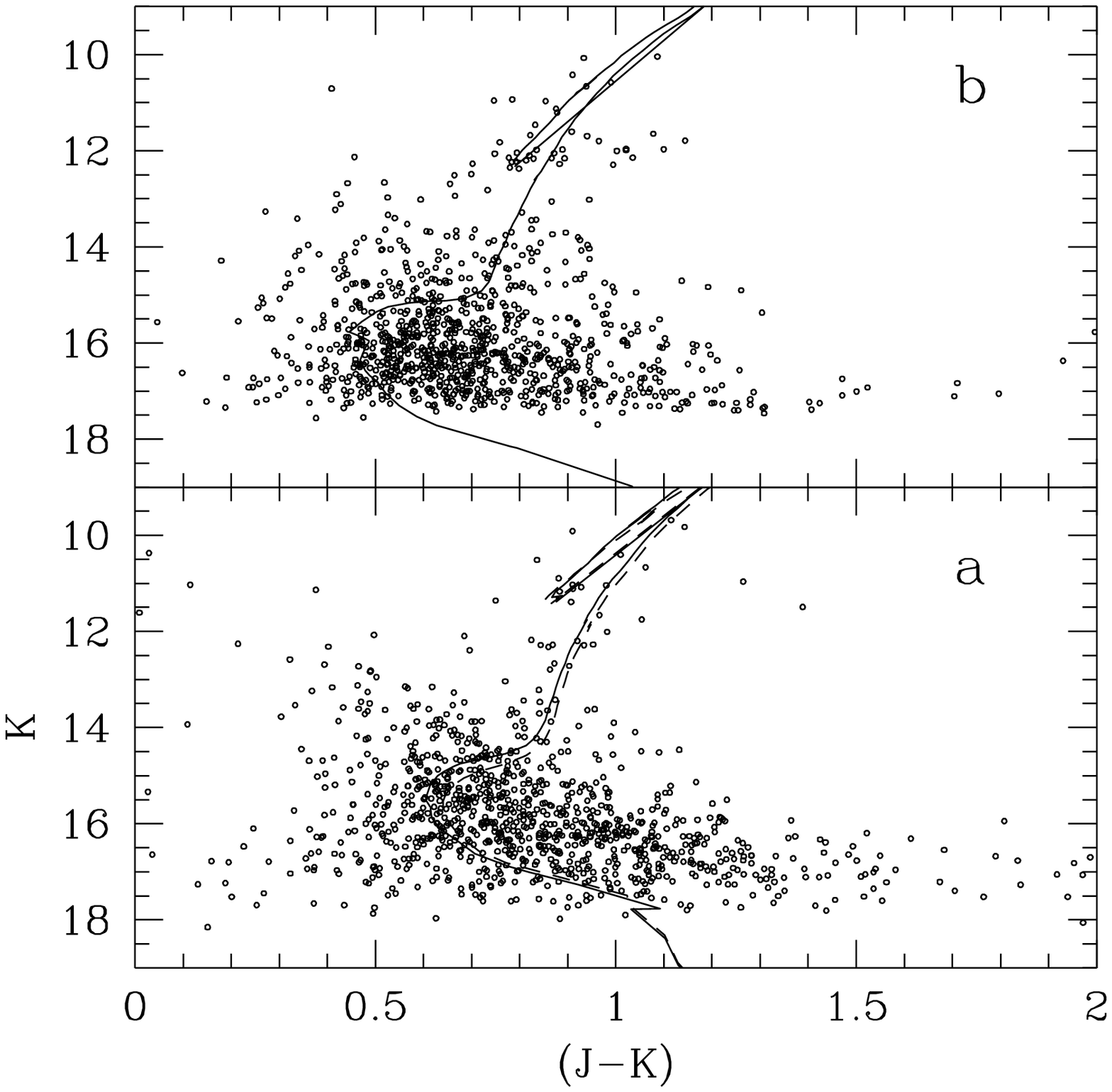,width=9cm,height=12cm}}
\caption{Comparison of clusters' CMD with theoretical isochrones.
Panel a) shows the CMD of Berkeley~17, with overimposed two $Z~=~0.008$
isochrones for the ages of $9~Gyr$ (solid line) and $13~Gyr$ (dashed
line). Reddening and distance modulus are $0.30~mag$ and $12.90~mag$,
respectively.
Panel b) shows the CMD of Berkeley~18, with overimposed a $Z~=~0.019$
isochrone of $4~Gyr$. Reddening and distance modulus are $0.25~mag$ and
$14.00~mag$,
respectively.}
\end{figure}

\noindent
Adopting the relation (Girardi et al 1998): 

\begin{equation}
[Fe/H] = log \frac{Z}{0.019}
\end{equation}

\noindent
we can derive a theoretical metal abundance $Z~\approx~0.008$. This value
is adopted to generate isochrones from the theoretical models by Girardi
et al (1998).
The best fit is show in Fig~5, panel a). It has been obtained overimposing
an $9~Gyr$ isochrone shifted by $(m-M)_{K}~=~12.90$ and $E(J-K)~=~0.30$.
These value are uncertain within $0.1~mag$.\\
From the relation 

\begin{equation}
\frac{E(J-K)}{E(B-V)} = 0.52
\end{equation}
 
\noindent
taken from Cardelli et al (1989), $E(J-K)$ transforms into
$E(B-V)=0.58~mag$.
As a consequence, $(m-M)_{o}$ becomes 12.00. This puts Berkeley~17  
$2.5~kpc$ away from the Sun. Reddening and distance are in agreement with
Phelps (1997) results.\\
As for the age, we find a value consistent with Kaluzny (1994)
suggestions, but significantly lower than Phelps (1997) estimate.
We believe that this difference 
arises from the poor fit of the clump stars in the Phelps (1997) study.
To cast light on this point, we overimposed (see Fig.~5, panel a))
to the cluster CMD a $13~Gyr$ isochrone. Fixing the clump color and
magnitude, the TO appears too faint, and the RGB too red. On the other
hand, trying to fix the color and magnitude of
the TO, which is more blurred, the RGB clump turns 
out to be bluer and brighter than observed. This is confirmed by Carraro
et al (1998), who find that Berkeley~17 is about $9~Gyr$ old analyzing
published $BVI$ photometry.

\subsection {The Luminosity Function of Berkeley~17}

To confirm the age determination based on the CMD,
 we compare the luminosity function (LF) of the MS of the
cluster (defined as
stars fainter than $K=15$ or bluer than $(J-K)=0.75$)
 with the simulations of two
populations of age $8~Gyr$ and $13~Gyr$ having $Z=0.008$.

First, by means of the usual experiments
 with artificial stars we derive the incompleteness correction $\gamma$ for Be 17, defined as the ratio of recovered on added stars.
 It turns out to be $\gamma= 1, 0.86, 0.20$ for $K < 15.5$, 16, $17~mag$.
The theoretical LFs are corrected for  incompleteness due to crowding, 
dividing  them by $\gamma$.

Second, we  derive the LF of the disc stars contaminating
the cluster using the model of the Galactic disc proposed by
Ng et al (1995) as modified by Vallenari et al
1998.
This model has been tested fitting CMD and luminosity functions of
low latitude Galactic fields (see Schmidtobreick et al 1998 for details).
Since this cluster is located at low Galactic latitude, the main contamination
is given by the thin disc.
Starting from a double exponential density distribution of the Galactic
thin disc, assuming a scale height of $250~pc$, 
a scale length of $2000~pc$, constant star formation rate
from $10~Gyr$ to $0.1~Gyr$,
we derive the simulated LF of the disc component, imposing the
number of stars observed in the range $12 <  K < 14$.
Changing the input parameters of the simulations (scale height and
scale length of the disc), we estimate that the
error on the disc LF at the level of the expected turnoff
of Berkeley~17 (i.e. between $14$ and $15~mag$) is about $10-15\%$.

Finally the observational LF of the MS
is obtained, subtracting the disc MS contamination
by the initial LF.
In Fig.~6 we compare  the simulated
LFs with the data. For magnitudes fainter
than $K \sim 16$ the incompleteness correction becomes relevant
and the comparison is not significant.
For brighter magnitudes, the oldest age gives a turnoff fainter
than the observed one, whereas the youngest one is in reasonable agreement
with the data. This result might be weakened by the fact that the LF of the
disc
population has not been derived from observations, but simulated.
However, to reproduce the data with an
age of 13 Gyr, we need to increase the  disc population of about $30\%$,
which is much higher than the estimated uncertainty.

As a conclusion, on the basis of the analysis of the CMD
and LF, although an age of $13~Gyr$ cannot be completely ruled
out, a much younger age ($\sim 8$ Gyr) seems to be favored.

\begin{figure*}
\centerline{\psfig{file=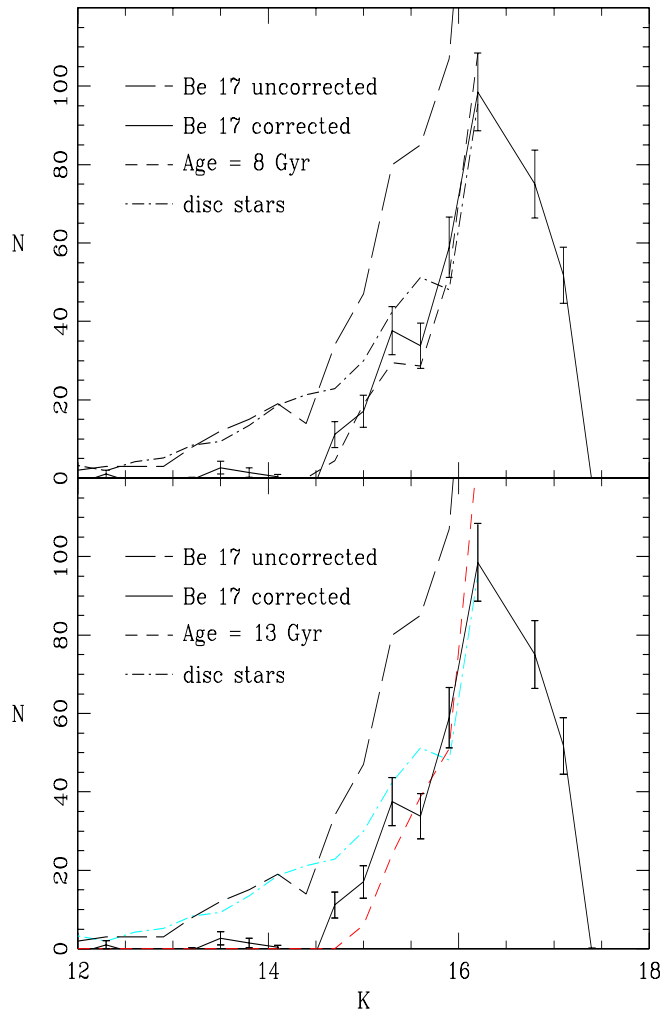,width=12cm,height=15cm}}
\caption{The LF of MS of Berkeley~17 is 
compared with theoretical simulations with ages of $8$ and $13~Gyr$
corrected for incompleteness. 
The disc population is simulated and subtracted from the observed LF.
The long dashed line represents the original data, dashed-dotted
the disc population, solid the MS LF of Berkeley~17
decontaminated by disc stars, and  finally short-dashed  the
simulation of the cluster for the labelled age,}
\end{figure*}

\section{Berkeley~18: cluster parameters}

\subsection{Metallicity}
No spectroscopic determinations of Berkeley~18 metallicity are available,
so we rely only on the slope of the RBG, and derive metallicity from eq.~(3).
We obtain $\Delta (J-K) / \Delta K = -0.125\pm0.05$, which translates in
solar metal abundance $[Fe/H] \approx 0.0$.

\subsection{Age, Distance and Reddening}
Berkeley~18 has been studied by Kaluzny (1997). From this preliminary work,
Kaluzny suggested that Berkeley~18 is as old as M~67 ($4~Gyr$, Carraro et
al 1996),
its reddening $E(B-V)$ is probably higher than $0.46~mag$, and the
distance from the Sun is around $5.8~kpc$.\\
In Fig.~5 (panel b) we present the CMD of Berkeley~18 with a best fitting 
isochrone of
$4~Gyr$ for the metal  abundance $Z~=~0.019$. 
From this fit we derive a reddening $E(J-K)~=~0.25~mag$ and an apparent
distance modulus $(m-M)_{K}~=~14.00~mag$.The errors in these estimates 
are around $0.1~mag$.\\
The corresponding visual reddening $E(B-V)$ turns out to be $0.40~mag$ 
from eq. (5), in agreement with Kaluzny's (1997) suggestions.
Using this value, we estimate the distance of Berkeley~18 to the Sun 
to be about $4.5~kpc$, significantly lower than Kaluzny's (1997)
estimate.

\begin{table}
\tabcolsep 0.10truecm
\caption{Derived parameters of the studied clusters.}
\begin{tabular}{lccccc}
\hline
\hline
\multicolumn{1}{c}{Cluster} &
\multicolumn{1}{c}{$[Fe/H]$} &
\multicolumn{1}{c}{$Age$} &
\multicolumn{1}{c}{$E(B-V)$} &
\multicolumn{1}{c}{$(m-M)_{o}$} &
\multicolumn{1}{c}{Distance}\\
 &dex&Gyr&mag&mag&kpc \\   
\hline
Berkeley~17    &  $\sim -0.35$ & $9\pm1$ & 0.58 & 12.00 & 2.5\\
Berkeley~18    &  $\sim  0.00$ & $4\pm1$ & 0.48 & 13.25 & 4.5\\
\hline\hline
\end{tabular}
\end{table}

\section{Conclusions}
We have presented and discussed near IR photometry of two poorly studied
old open clusters, Berkeley~17 and Berkeley~18.
From the analysis of the CMDs we derived cluster fundamental parameters
which are listed in Table~3. In details, we derived a photometric estimate
of the cluster metal abundance (column~2), new age estimates with
uncertainties
derived from the fit of several isochrones of different ages (column~3), 
reddening (column~4), distance modulus (column~5) and heliocentric
distance (last column).\\ 
The main results of this study can be summarized
as follows.

\begin{description}
\item $\bullet$ using the calibration between the slope of the RGB and the metallicity (Tiede et al
1997), we estimate that
Berkeley~17 has a metal abundance $[Fe/H]~\sim~-0.35$, while Berkeley~18
has about solar metal abundance;
\item $\bullet$ Berkeley~17 is found to be $9~Gyr$ old. The difference
with Phelps (1997) age estimate (about $12~Gyr$)is probably due to a
better fitting of the
clump stars, which are more clearly visible in our study.
This younger age is also suggested  by the analysis of the cluster LF
corrected for the disc star contamination.\\
 This new age determination implies  that Berkeley~17 
is no longer the oldest known open cluster, and suggests the possibility
that the Milky Way experienced a minimum of star formation between 10 and
13 $Gyr$ ago;
\item $\bullet$ Berkeley~18 is as old as M~67, but its distance is 
significantly lower than previous estimates.
\end{description}

\begin{acknowledgements}
We thank L. Hunt and F. Mannucci for kind assistance
during data reduction. 
This study has been financed by the Italian Ministry of
University, Scientific Research and Technology (MURST) and the Italian
Space Agency (ASI).
The work of L\'eo Girardi is funded by the Alexander von
Humboldt-Stiftung.

\end{acknowledgements}

{}

\end{document}